%% file: Optimal-Design-layout.tex
\documentclass[VANCOUVER]{USG} 
\usepackage{anyfontsize} %

\usepackage{colortbl}   
\usepackage{xcolor}     

\usepackage{steinmetz}
\graphicspath{{./images/}}

\newcommand{\fsv}{FSV} 
\newcommand{\nn}{NucNet} 

\articletype{Research Article}%

\received{}
\revised{}
\accepted{}
\journal{}
\volume{}
\copyyear{}
\articledoi{}

\newcommand{\super}[1]{$^{\textrm{\tnote{#1}}}$} 

\begin{document}
\title{7 Tesla Quantitative MRI and Machine Learning for Exploratory Motor Subtype Stratification and Diagnosis in Parkinson’s Disease}
\author[1,2]{Anne Louise Kristoffersen, MSc}
\author[2,3]{Runa Geirmundsdatter Unsgård, MD}
\author[1]{Marc-Antoine Fortin, MSc}
\author[4]{Ingrid Gylterud Kvålsgard}
\author[4,5]{Kjerst Eline Stige, MD}
\author[4,5]{Thanh Pierre Doan, MD, PhD}
\author[2,3]{Erik Magnus Berntsen, MD, PhD}
\author[6,7,8,9]{Charalampos Tzoulis, MD, PhD}
\author[1,2]{Pål Erik Goa, PhD}

\address[1]{\orgdiv{Department of Physics, }\orgname{Norwegian University of Science and Technology, }%
\orgaddress{\state{Trondheim, }\country{Norway}}}
\address[2]{\orgdiv{Department of Radiology and Nuclear Medicine, }\orgname{St Olav’s University Hospital, }%
\orgaddress{\state{Trondheim, }\country{Norway}}}
\address[3]{\orgdiv{Department of Circulation
and Medical Imaging, }\orgname{Norwegian University of Science and Technology, }%
\orgaddress{\state{Trondheim, }\country{Norway}}}
\address[4]{\orgdiv{Department of Neuromedicine and Movement Sciences, }\orgname{Norwegian University of Science and Technology, }%
\orgaddress{\state{Trondheim, }\country{Norway}}}
\address[5]{\orgdiv{Department of Neurology and Clinical Neurophysiology, }\orgname{St Olav’s University Hospital, }%
\orgaddress{\state{Trondheim, }\country{Norway}}}
\address[6]{\orgdiv{Department of Neurology, }\orgname{Haukeland University Hospital, }%
\orgaddress{\state{Bergen, }\country{Norway}}}
\address[7]{\orgdiv{Neuro-SysMed Center for Clinical Treatment Research, Department of Neurology, }\orgname{Haukeland University Hospital, }%
\orgaddress{\state{5021 Bergen, }\country{Norway}}}
\address[8]{\orgdiv{Department of Clinical Medicine, }\orgname{University of Bergen, }%
\orgaddress{\state{Pb 7804, 5020 Bergen, }\country{Norway}}}
\address[9]{\orgdiv{K.G. Jebsen Center for Translational Research in Parkinson’s disease, }\orgname{University of Bergen, }%
\orgaddress{\state{Pb 7804, 5020 Bergen, }\country{Norway}}}

\shrttitle{7T qMRI and ML for PD Stratification}

\keywords{Parkinson's Disease | qMRI | classification | machine learning | high-field MRI}

\titleack{ChatGPT and Microsoft Copilot were used for language proofreading and suggesting Python libraries and debugging. AI was not used to generate ideas or data.}

\abstract[ABSTRACT]{BACKGROUND:
Parkinson’s disease (PD) is a highly heterogeneous disease, including which motor symptoms are dominating. Imaging biomarkers that support subtype stratification could also improve biological understanding and study design, and enable personalized treatment strategies.

PURPOSE:
Evaluate whether deep-learning based automatic brain segmentation, in addition to quantitative maps from 7 Tesla MRI, can highlight differences between Healthy Controls (HC), Postural Instability and Gait Difficulty (PIGD) and Tremor Dominant (TD), and subsequently be used for objective PD stratification. The performance of machine learning classifiers may be improved with feature selection.

STUDY TYPE:
Retrospective case–control study.

SUBJECTS:
21 HC, and 24 people with PD (PwP) (14 PIGD, 9 TD; 1 indeterminate - excluded from subtype analyses).

FIELD STRENGTH/SEQUENCE:
7T MRI using the MP2RAGE, ASPIRE sequences.

ASSESSMENT:
The U-Net training was assessed with Dice Sørensen Coefficient (DSC).
Two classification approaches using five‑fold cross‑validation were defined across three tasks: (1) HC vs PwP; (2) PIGD vs TD; (3) multiclass, HC vs PIGD vs TD. Approach A used all extracted features. Approach B found the optimal subset of features for the classification tasks.

STATISTICAL TESTS:
Accuracy and AUC reported per fold and averaged across the five folds for evaluating classification performance. One‑vs‑rest area under the ROC curve (AUC) for multiclass tasks.

RESULTS:
The U-Net achieved mean DSC of 0.86 for all ROIs during training.
Approach A: Task 1 best accuracy of 0.69 and best AUC of 0.73. Task 2 accuracy 0.69, AUC 0.90. Task 3 accuracy 0.62, AUC 0.66.
Approach B: 
Task 1 accuracy of 0.82 and AUC of 0.93. Task 2 accuracy 1.00, AUC 1.00. Task 3 accuracy 0.73, AUC 0.91.

DATA CONCLUSION:
DL‑based segmentation combined with qMRI feature selection improved classification relative to using all features, supporting the potential of interpretable, low‑dimensional imaging signatures for PD diagnosis support and phenotype stratification. Larger, multi‑site studies are warranted to assess generalizability and stability.

}







\maketitle

\include{sections/introduction}

\include{sections/methods}

\include{sections/results}

\include{sections/discussion}


\bmsubsection*{Acknowledgments}
ChatGPT and Microsoft Copilot were used for language proofreading and suggesting Python libraries and debugging. AI was not used to generate ideas or data.

\bmsubsection*{Financial Disclosure}

None reported.

\bmsubsection*{Conflicts of Interest}

The authors declare no conflicts of interest.

\bibliography{sections/references_230326}












\newpage
\include{sections/supplementary}
\end{document}

%% file: sections/introduction.tex
\section{Introduction}\label{sec1}
Parkinson's disease (PD) is the fastest growing disorder, and is currently the second most common neurodegenerative disease, \citep{schneider_clinical_2015}. Neurodegeneration across the central and autonomic nervous systems gives rise to a multitude of progressive motor and non-motor symptoms. Primary motor symptoms include tremor, rigidity, and bradykinesia, while non-motor symptoms include sleep disorder, psychiatric symptoms, gastrointestinal dysfunction, and cognitive decline \citep{lees_parkinsons_2009, dorsey_emerging_2018}.
The disease is highly heterogeneous in terms of symptoms, progression rate, age of onset, and treatment response \citep{greenland_clinical_2019,chen-plotkin_finding_2018}.

PD diagnosis is currently primarily based on clinical criteria, with the main role of MRI being to exclude differential diagnoses \citep{mahlknecht_significance_2010, heim_magnetic_2017}. While conventional MRI is mainly used to exclude mimics, quantitative MRI (qMRI) has increasing potential to provide objective markers of tissue properties relevant to PD pathophysiology. Current treatments are purely symptomatic, providing some transient relief mainly for motor symptoms, rather than disease-modifying therapies to delay disease progression \citep{bloem_parkinsons_2021}.

There is currently an increased focus on stratification of PD, especially in the context of personalised medicine. PD is most commonly classified according to motor phenotype into postural instability and gait difficulty (PIGD) or tremor dominant (TD) subtypes \citep{thenganatt_parkinson_2014}, associated with different disease trajectories. PIGD is commonly associated with gait disability, higher incidence of dementia, and faster functional decline, whereas TD tends to progress more slowly \citep{kalia_parkinsons_2015}. However, phenotype assignment can be rater- and context-dependent and may fluctuate with medication state. An objective imaging-derived marker could therefore standardise stratification across cohorts and support trial enrichment, and ultra-high-field qMRI may be particularly useful as a discovery platform because motor-circuit nuclei are small and sensitive to partial-volume effects.

Recent studies have demonstrated the feasibility of differentiating PD from healthy controls (HC) using neuroimaging, with high level of accuracy \citep{desai_enhancing_2024, zubair_classification_2024}.
Beyond diagnosis, several studies investigated group differences between PIGD and TD. Changes in myelin content and iron distribution in the basal ganglia have been detected \citep{cheng_synthetic_2025, zhang_distribution_2023, li_quantitative_2018, thomas_brain_2020}.
Other studies extended beyond investigating quantitative group differences or primarily diagnosing PD, and achieved high classification performance. Various subtype classification has been done using resting-state fMRI, DTI, myelin content, T1 and T2 \citep{gu_automatic_2016, hosseini_cross-regional_2025, cheng_explainable_2025}.

These studies imply that structural and functional MRI contain valuable information for characterising individuals as HC or with a PD subtype. Although structural MRI was usually used in addition to functional imaging, some of the mentioned studies showed that structural imaging contains sufficient information. Additionally, it is important to note that the functional MRI measures are highly sensitive to the patient’s dopaminergic state. The aim of this study was to investigate whether qMRI derived from structural images can support diagnosis and motor phenotype stratification, without the need for functional MRI. We treat 7T as a discovery platform that enables higher contrast and improved delineation of small subcortical nuclei; the goal is to identify candidate qMRI signatures that can later be evaluated for robustness and portability at clinical field strengths.

%% file: sections/methods.tex
\section{Methods}
All participants provided written informed consent prior to inclusion. The Regional Committee for Medical and Health Research Ethics have approved both the STRAT-PARK (REK 74985) and the 7 Tesla study (REK 108066).

Two automatic segmentation tools; FastSurferVINN (FSV) and an in-house trained U-Net were applied to 7 Tesla MRI data. Features were extracted from Quantitative Susceptibility Mapping (QSM), R1 and R2* maps within the automatically segmented regions of interest (ROIs). These features served as input for machine learning (ML) classifiers with the goal of assessing their classification performance in \textit{three tasks}, using \textit{two approaches}.

\subsection{Participants}
The participants in this study were a subset of the STRAT-PARK cohort study \cite{stige_strat-park_2024}, specifically those who underwent 7T MRI scanning in Trondheim. To increase the number of HC, a subset from a 7 Tesla study on HC scanned with the same scanner and sequence was included. A total of 45 individuals were included: 24 PwP, (11 females, 13 males, mean age 65.8 years (SD=6.7)), and 21 age-matched HC (11 females, 10 males, mean age 60.2 (SD=9.2)). Of the 24 PwP 14 had the motor phenotype PIGD, 9 had TD and 1 had indeterminate. Motor phenotype was classified as PIGD or TD using the MDS-UPDRS TD/PIGD score following the algorithm in \cite{stebbins_how_2013}. Ratings were mainly obtained in ON medication state. Phenotype labels were calculated prior to qMRI analysis. In the classification tasks where motor phenotype is predicted, the indeterminate individual is removed, resulting in 23 PwP.

A second cohort of 15 young healthy adults (mean age of 25.9 years) were also imaged as part of the previously mentioned 7 Tesla study. These images were used to train the U-Net, NucNet, as described in Section \ref{sec:nnunet-methods}. 

\subsection{MR Imaging}\label{sec:MRI-methods}
All images were acquired using a Siemens Magnetom Terra 7T MR system (Siemens Healthineers, Erlangen, Germany) equipped with a 1Tx/32Rx head coil (Nova Medical Inc, Wilmington MA). The protocol included an MP2RAGE sequence \citep{marques_mp2rage_2010} and a multi-echo GRE sequence (ASPIRE) \citep{eckstein_computationally_2018}. The MP2RAGE sequence had a repetition time (TR) of 4300 ms, echo time (TE) of 1.99 ms, first and second inversion times of 840 and 2370 ms, first and second flip angles (FA) of 5.0 and 6.0 deg, and a resolution of (0.75mm)$^{3}$ isotropic. The multi-echo GRE used TR of 31.0 ms, four echos of TE 2.54, 7.22, 14.44 and 23.23 ms, FA of 12 deg and resolution of (0.75mm)$^{3}$ isotropic. The Total Generalized Variation (TGV) pipeline was used to derive QSM from the multi-echo GRE sequence \citep{langkammer_fast_2015}. 

\begin{figure}
    \centering
    \includegraphics[width=0.8\linewidth]{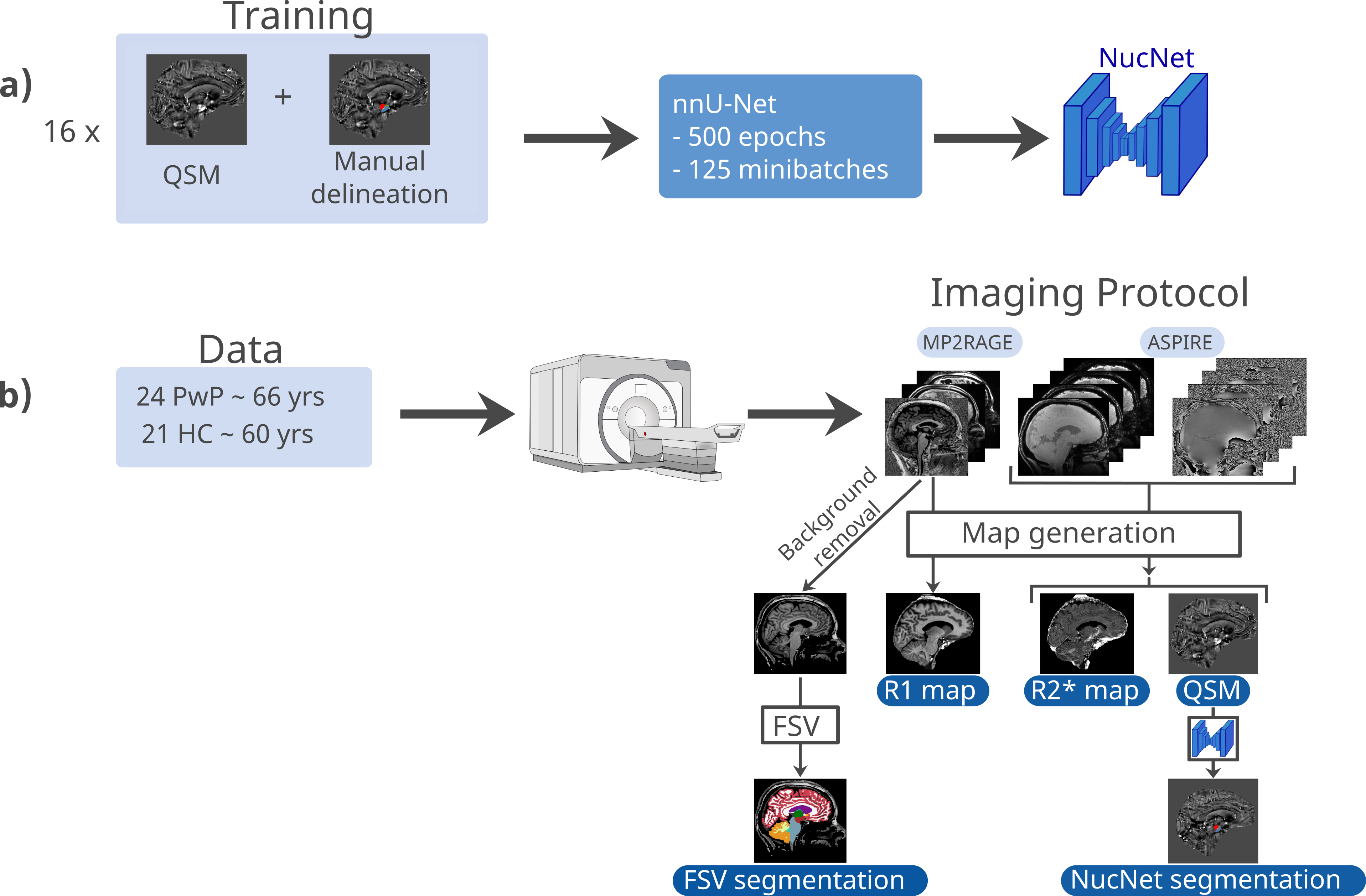}
    \caption{a) Training a 3D U-Net using the nnU-Net framework. The trained U-Net, \nn{}, segments three labels: substantia nigra (SN), red nucleus (RN) and subthalamic nucleus (STN) on QSM images. b) Pipeline for obtaining the quantitative maps and segmentation of the ROIs.}
    \label{fig:fig1}
\end{figure}

QSM is inherently quantitative, but quantitative maps from the T1-weighted (T1w) and T2*-weighted (T2*w) images were also required in order to utilise the images for qMRI.
Using the \texttt{nighres.intensity} module from Nighres \citep{huntenburg_nighres_2018}, R1 maps were generated from the MP2RAGE images and the R2* maps from ASPIRE.
This pipeline is illustrated in Figure \ref{fig:fig1}b.

\subsection{Feature Extraction}\label{sec:feature_extraction-methods}
All features were extracted from ROIs obtained by automatic segmentation, which were visually inspected for gross errors. None were excluded.
Two automatic segmentation tools were used. Several ROIs important for PD have inadequate contrast on T1w images, which most tools are designed for. Consequently, a U-Net, \nn{}, was trained to segment substantia nigra (SN), red nucleus (RN), and subthalamic nucleus (STN) on QSM. The rest of the ROIs used in this study were obtained using \fsv{} on MP2RAGE. 

\subsubsection{FSV}
The deep-learning based tool FSV segments the whole brain into 95 classes, mimicking the atlas-based tool FreeSurfer's anatomical segmentation and cortical parcellation (DKTatlas) \citep{henschel_fastsurfer_2020, henschel_fastsurfervinn_2022}.
The tool is optimised for segmenting T1w images acquired at 1.5T and 3T. In order to use the MP2RAGE images with FSV, the background noise was removed using the Otsu thresholding method from scikit-image package \citep{walt_scikit-image_2014}. The \fsv{} call was executed with the flags \texttt{--device cpu}, \texttt{--seg\_only} and \texttt{--vox\_size 0.75}. 

Out of the 33 subcortical classes, only the ROIs exhibiting the most clearly defined anatomical boundaries upon visual inspection were kept. As this study is inspired by a data-driven approach, this selection aimed to maximise the number of ROIs included while maintaining high-quality ROIs. This was based on prior experience during visual assessment. The lateral ventricles, thalamus, caudate, putamen, hippocampus and amygdala were included in this study. 

\subsubsection{\nn{}}\label{sec:nnunet-methods}
To overcome the problem of ROIs with poor contrast in T1w images, a 3D U-Net, \nn{}, was trained to segment QSM images using the nnU-Net framework (v2.1.1) \cite{isensee_nnu-net_2021}. SN, RN and STN lack contrast on T1w images but have good contrast on QSM, which was used as the contrast for training \nn{}. 

\begin{figure}
    \centering
    \includegraphics[width=0.8\linewidth]{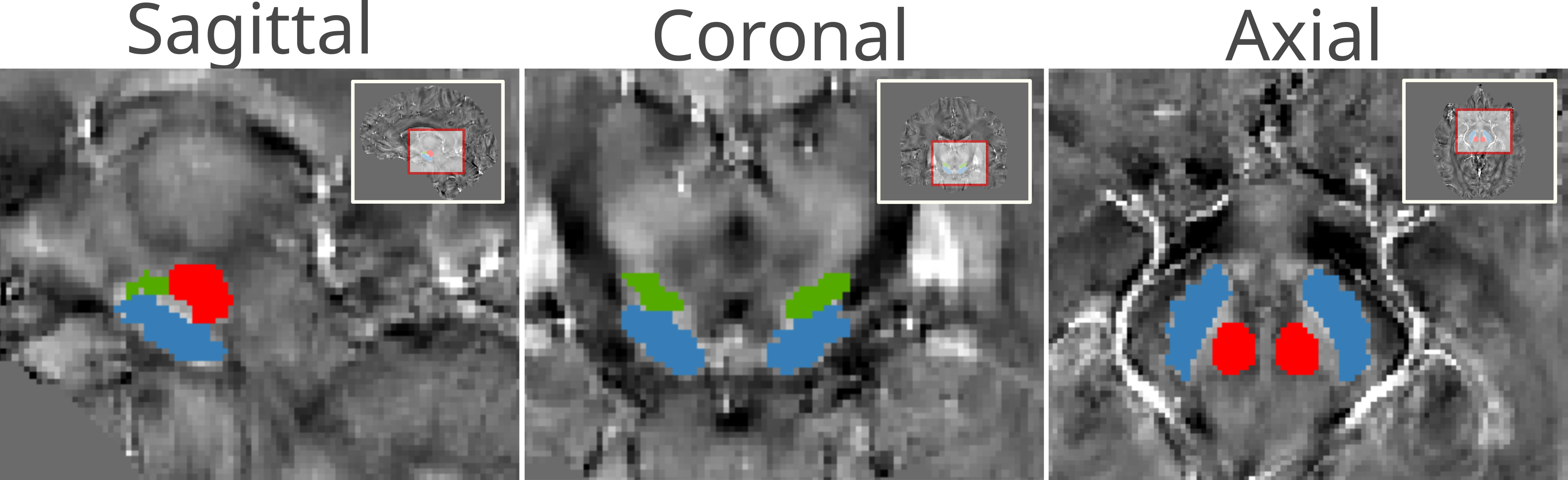}
    \caption{Example of manual delineation of substantia nigra (SN), red nucleus (RN) and subthalamic nucleus (STN) on QSM used for training \nn{}. SN is shown in blue, RN in red and STN in green. In a sagittal, coronal, and axial view. All images are zoomed in, and location is indicated in the top right corner in all.}
    \label{fig:fig2}
\end{figure}

The training data set consisted of 16 manual delineations by two experts, 15 from young healthy adults (mean age of 28 years) and one PwP. Expert 1 (**) delineated the MRIs from the 15 young healthy adults, and is a trained radiologist. Expert 2 (**) delineated the one PwP as part of another project, and is a trained neuroanatomist. The 15 young adults are from a different cohort than the one used for the data generation in this study, however the one PwP comes from STRAT-PARK. Figure \ref{fig:fig2} shows an example of a manual delineation of SN, RN and STN used for training. The individual is a young healthy adult, not part of the individuals used in the classification tasks.

The 3D full resolution U-Net configuration was used, with the default loss function and optimiser. The number of epochs was decreased to 500, and the number of mini-batches was decreased to 125 after inspection from the early trainings. Five-fold cross-validation was used to mitigate overfitting, and the standard nnU-Net data augmentation was used to increase the training data pool. 

To extract the quantitative values in all maps, the ROIs were co-registered to the other contrasts using the \texttt{nighres.registration} module from Nighres \citep{huntenburg_nighres_2018}.

\subsection{Data Generation}
The pipeline, from acquiring MR images to final segmentations and quantitative maps, is illustrated in Figure \ref{fig:fig1}. 
The pipeline of training the U-Net, \nn{}, is illustrated in Figure \ref{fig:fig1}a, and is explained in further detail in Section \ref{sec:nnunet-methods}. In Figure \ref{fig:fig1}b, the creation of quantitative maps and segmentations is shown, and further explained in Sections \ref{sec:MRI-methods} and \ref{sec:feature_extraction-methods}.

\subsubsection{Subject Vector}
For each individual, a subject vector was created. The first set of features in the vector was the volume of each of the ROIs. The volumes were normalised on total intracranial volume (TIV) using SPM12 \citep{noauthor_spm12_nodate}: \textit{SPM12 (7771) $\rightarrow$ Spatial $\rightarrow$ Segmentation} to obtain the grey matter, white matter and cerebrospinal fluid probability masks. To obtain the tissue volumes comprising the TIV \textit{SPM12 $\rightarrow$ Utils $\rightarrow$ Tissue Volumes} was ran.

For each of the nine ROIs, eight statistical metrics from the map values from the three quantitative maps were extracted. The nine ROIs selected were: SN, RN, STN, lateral ventricles, thalamus, caudate, putamen, hippocampus and amygdala. The eight statistical metrics were: mean, standard deviation, median, maximum value, minimum value, 5-percentile, skewness and kurtosis. The quantitative values were extracted from the maps: R1 map, R2* map and QSM. The total subject vector had 225 features: 9 ROI volumes + (3 contrasts x 9 ROIs x 8 statistical metrics). 

To prevent bias towards the scale of the feature values in the classification tasks, z-score normalisation was applied to normalise the features.

\subsection{Classification}
Three classification tasks were evaluated, each with two different approaches. Task 1 evaluated how well HC and PwP could be differentiated based on the features in this study. Task 2 was classifying PwP into the two subgroups based on motor phenotype: PIGD or TD. Classification Task 3 went on to classify the individuals into three classes: HC, PIGD and TD. 

The two approaches were based on different number of features. Approach A used all features, and tested a variety of different ML classifiers. Approach B employed Support Vector Machine (SVM), and identified and then used the optimal subset of features. 

The training and analyses were performed in Python 3 using ML classifiers from the packages \texttt{scikit-learn} \citep{pedregosa_scikit-learn_2011} and \texttt{xgboost} \citep{chen_xgboost_2016}. For both approaches, cross-validation with five folds was used and reported. 

\subsubsection{Approach A: All Features}
Five ML classifiers were tested: SVM, random forest, logistic regression (LogReg), XGBoost and K-Nearest-Neighbours (KNN). All classifiers were tuned using a grid search tailored for the specific classifier. During the grid search, the model was refitted using the configuration that achieved the highest accuracy on the validation set.

The hyperparameters tuned for LogReg were the regularisation parameter and the penalty. For SVM, the regularisation parameter, kernel, gamma for the relevant kernels, and degree for the relevant kernels, were tuned. For Random Forest, the number of trees, maximum depth of the trees, minimum number of samples to split internal node, and minimum number of samples required to be a leaf node were optimised. For XGBoost the number of trees, maximum depth of the trees, learning rate, subsample ratio of training instances and subsample ratio of columns were tuned. For KNN the number of neighbours, the weight function in prediction and the metric to compute distance were optimised.

For certain classifiers, the weighting of the features can be extracted and subsequently used to determine the importance of the features in the classification process. This applies to LogReg, Random Forest and XGBoost. These were also reported.

\subsubsection{Approach B: Optimal Subset of Features}\label{sec:methods_optimal-number-of-features}
For all individual features, the area under the receiver-operating curve (AUC) was calculated (one-vs-rest AUC for the three classes task). The 12 features yielding the highest individual AUC were noted down, to constitute a new smaller feature pool. For the two first tasks (HC vs PwP and PIGD vs TD), these pools were different and found independently of each other. For the third task, the top six features from each pool were combined into the third 12-feature-pool. In this approach, SVM (linear kernel) was used, due to its widespread use in related research \citep{zhang_mining_2022}.

From the 12 features found as explained above, the optimal subset of them was found as follows. The combination was determined by obtaining the fold-average accuracy and AUC for each combination of features (number of and set of features), which was found through a search through all the possible models from these 12 features. The workflow is illustrated in Supplementary Materials, Figure S1.

%% file: sections/results.tex
\section{Results}
\subsection{Segmentation results}
\subsubsection{\nn{} training}
Training \nn{} gave an overall dice score of 0.86 for the three ROIs. For the five folds, SN achieved an average dice score 0.88, for RN it was 0.93 and for STN it was 0.77. In a separate project, Expert 2 delineated these ROIs for 9 additional individuals included in the present study. Testing \nn{} against those gave an average dice score of 0.89 for SN, 0.93 for RN and 0.82 for STN.

\subsubsection{Segmentations}
\begin{figure}
    \centering
    \includegraphics[width=0.8\linewidth]{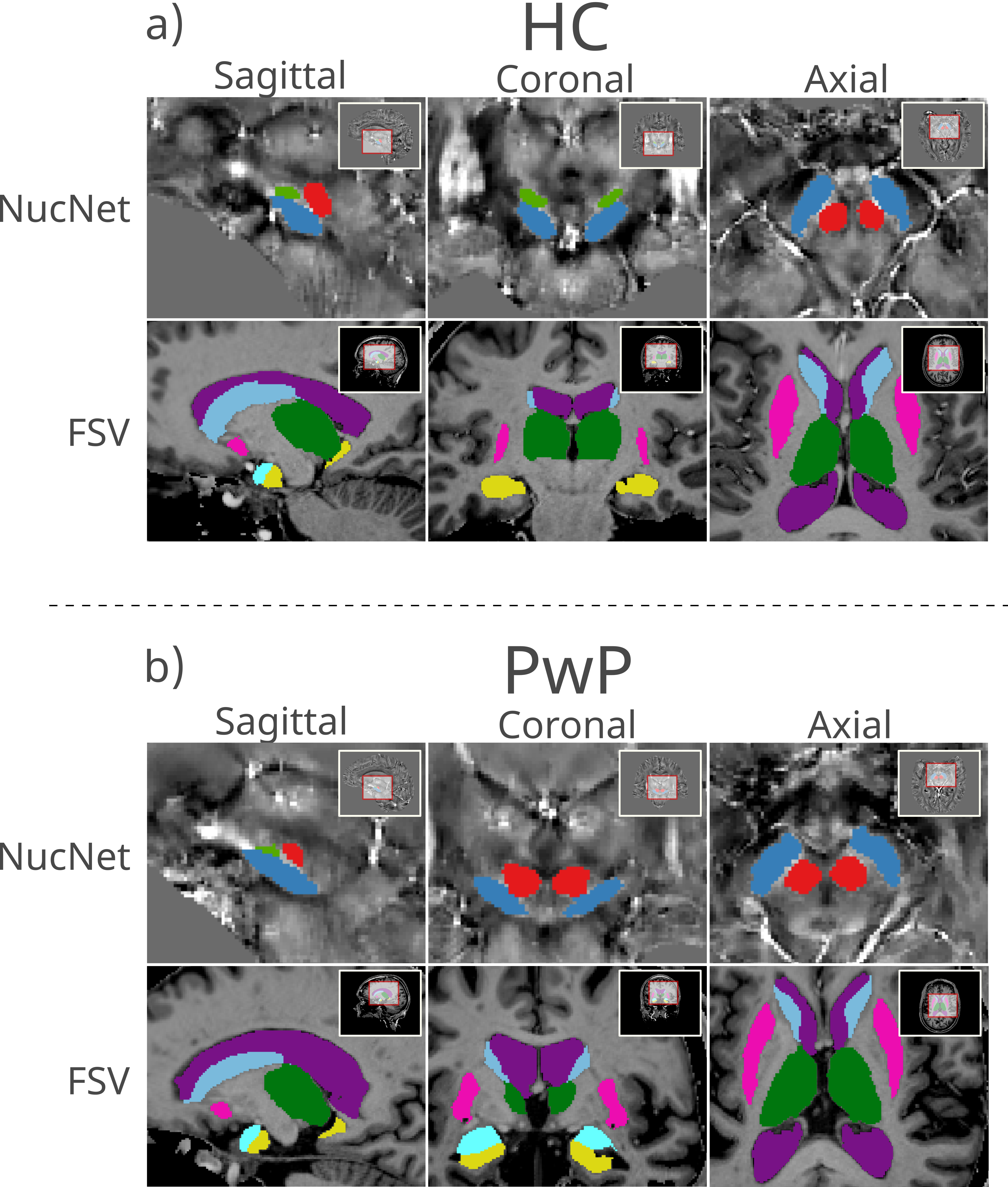}
    \caption{Segmentations shown for a) Healthy Control (HC) and b) Person with Parkinson's Disease (PwP). Top row shows an example of \nn{} segmentation of substantia nigra (SN), red nucleus (RN) and subthalamic nucleus (STN) in a sagittal, coronal plane and axial view. Bottom row shows the FastSurferVINN (FSV) segmentation from the same individual, of lateral ventricle is shown in purple, thalamus in green, caudate in light blue, putamen in pink, hippocampus in yellow and amygdala in turquoise in a sagittal plane, coronal plane and axial view. All images are zoomed in, and location is indicated in the top right corner in all.}
    \label{fig:fig3}
\end{figure}

Figure \ref{fig:fig3} shows examples of the segmentations for a) HC and b) PwP. Slices are not aligned because the ROIs from the segmentation tools are located in different positions in the brain. The figures shows the segmentation from \nn{} and \fsv{} in the sagittal, coronal and axial plane. The individuals are unseen for \nn{}.

In Table \ref{tab:mean-std-mapvalues-rois} the average mean and standard deviation of R1, R2 and QSM values for each ROI are presented.
\input{tables-figures/tabell1}

\subsection{Classification - HC vs PwP}
\subsubsection{Approach A}\label{sec:hcvspwp_all}
Despite evaluating a variety of classifiers, none of them classify individuals as HC or PwP in a satisfactory manner. LogReg achieved a slightly higher accuracy than the other classifiers, at 0.69. The classifier that achieved the highest AUC was also LogReg, at 0.73. 
In the left part of Table \ref{tab:all_classifiers-all_tasks} the accuracy and AUC of all the classifiers are listed. 

In Table \ref{tab:hcvspwp_features} the top seven features of LogReg, Random Forest and XGBoost are presented in the first five columns.

In Figure \ref{fig:fig4} the ROC curve for LogReg averaged over the five folds is shown in blue. In Supplementary Materials, Figure S2 shows the ROC curves of all the classifiers (as five-fold cross-validation was applied, there are five ROC curves in each plot, in addition to a mean ROC curve).

\subsubsection{Approach B}\label{sec:hcvspwp_optimal}
The 12 features that made up the smaller feature pool from which the optimal subset was identified are listed in the Supplementary Materials S.3.1. The optimal subset of features, yielding the highest AUC, was determined to consist of seven features.
The optimal subset of features is listed in the rightmost column in Table \ref{tab:hcvspwp_features}.
This yielded an accuracy of 0.82, and AUC of 0.93. 

In Figure \ref{fig:fig5} the ROC curve for the optimal subset averaged over the five folds, is shown in blue. In Supplementary Materials Figure S5 shows the ROC curves of all five folds.
\input{tables-figures/tabell2}

\subsection{Classification - PIGD vs TD}
\subsubsection{Approach A}\label{sec:pigdvstd_all}
For PIGD vs TD, the classifier that achieved the highest AUC for the \textit{all-features} method was SVM, at 0.90. LogReg achieved higher accuracy than the other classifiers, at 0.69. In the middle part of Table \ref{tab:all_classifiers-all_tasks} the accuracy and AUC of all classifiers are presented. 

In Table \ref{tab:pigdvstd_features} the top four features of LogReg, Random Forest and XGBoost are presented in the first five columns.

In Figure \ref{fig:fig4} the ROC curve for SVM averaged over the five folds is shown in red. In Supplementary Materials, Figure S3 shows the ROC curves of all the classifiers. As five-fold cross-validation was applied, there are five ROC curves in each plot, in addition to a mean ROC curve.

\subsubsection{Approach B}\label{sec:pigdvstd_optimal}
The 12 features that made up the smaller feature pool from which the optimal subset was identified are listed in the Supplementary Materials S.3.2. The combination of features that gave the highest AUC was determined to be four. The optimal subset of features is listed in the rightmost column in Table \ref{tab:pigdvstd_features}.
This yielded an accuracy of 1.00, and AUC of 1.00. 

In Figure \ref{fig:fig5} the ROC curve for the optimal subset, averaged over the five folds, is shown in red. In Supplementary Materials Figure S6 shows the ROC curves of all the five folds.
\input{tables-figures/tabell3}

\subsection{Classification - HC vs PIGD vs TD}
\subsubsection{Approach A}\label{sec:hcvspigdvstd_all}
For classification of HC vs PIGD vs TD, KNN achieved higher accuracy than the other classifiers, at 0.62. SVM achieved the highest AUC at 0.66. In the right part of Table \ref{tab:all_classifiers-all_tasks} the accuracy and AUC of all classifiers are presented. 

In Table \ref{tab:hcvspigdvstd_features} the top six features of LogReg, Random Forest and XGBoost are presented in the first five columns.

In Figure \ref{fig:fig4} the ROC curve for SVM averaged over the five folds is shown in green. In Supplementary Materials, Figure S4 shows the ROC curves of all the classifiers (as five-fold cross-validation was applied, there are five ROC curves in each plot, in addition to a mean ROC curve). 
\input{tables-figures/tabell4}

\subsubsection{Approach B}\label{sec:hcvspigdvstd_optimal}
The 12 features that made up the smaller feature pool for this task were the top six from HC vs PwP and the top six from PIGD vs TD. The combination of features that gave the highest AUC was determined to be 6. The optimal subset of features is listed in the rightmost column in Table \ref{tab:hcvspigdvstd_features}
This yielded an accuracy of 0.73 and AUC of 0.91. 

In Figure \ref{fig:fig5} the ROC curve for the optimal subset averaged over the five folds, is shown in green. In Supplementary Materials Figure S7 shows the ROC curves of all five folds.

\begin{figure}
    \centering
    \includegraphics[width=0.8\linewidth]{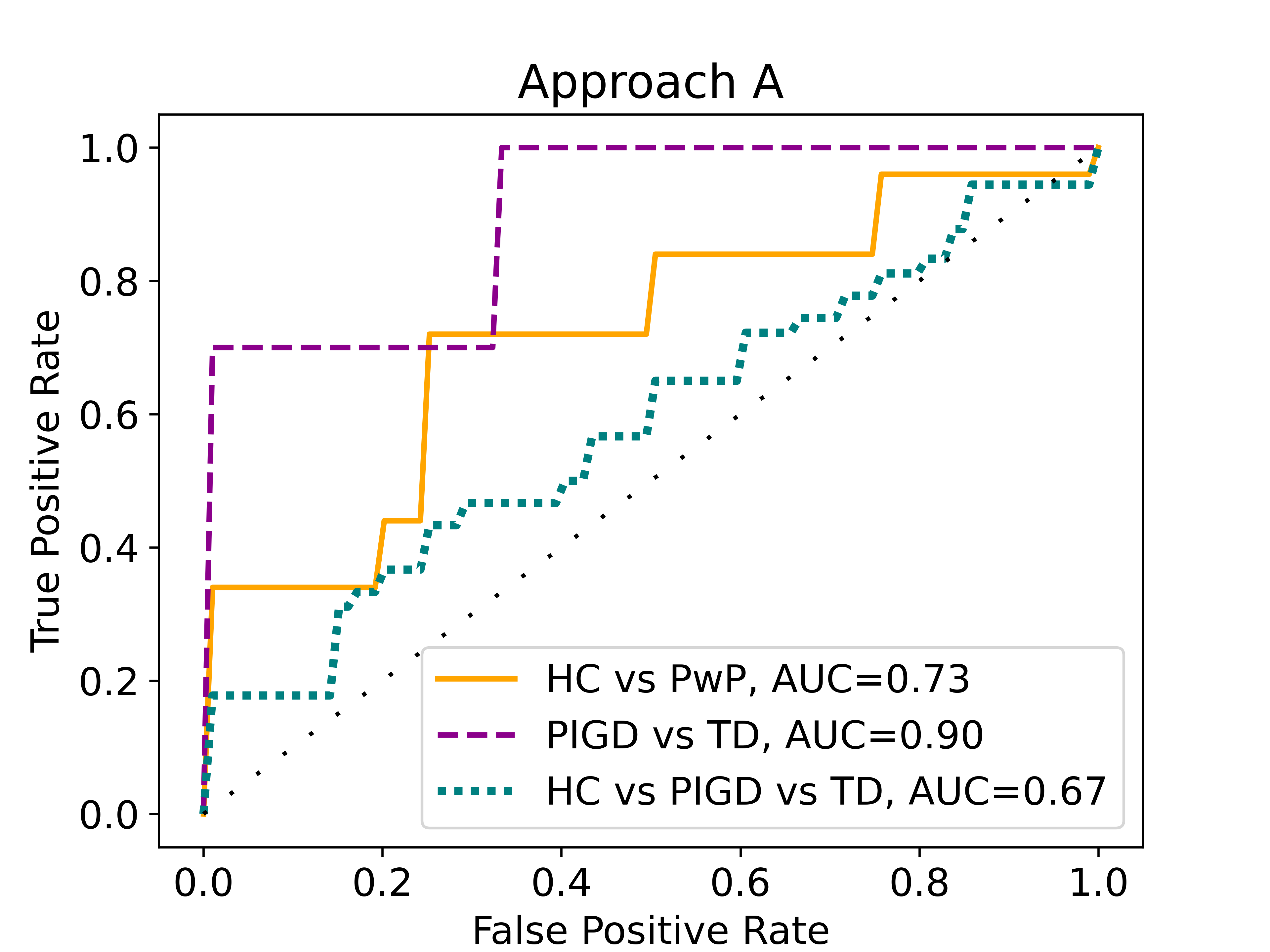}
    \caption{Approach A, using all features. ROC for Healthy Controls (HC) vs People with Parkinson's disease (PwP) in solid orange. ROC for Postural Instability and Gait Difficulty (PIGD) vs Tremor Dominant (TD) in dashed purple. ROC for HC vs PIGD vs TD in dotted green. Chance level is shown in loosely dashed black.}
    \label{fig:fig4}
\end{figure}

\begin{figure}
    \centering
    \includegraphics[width=0.8\linewidth]{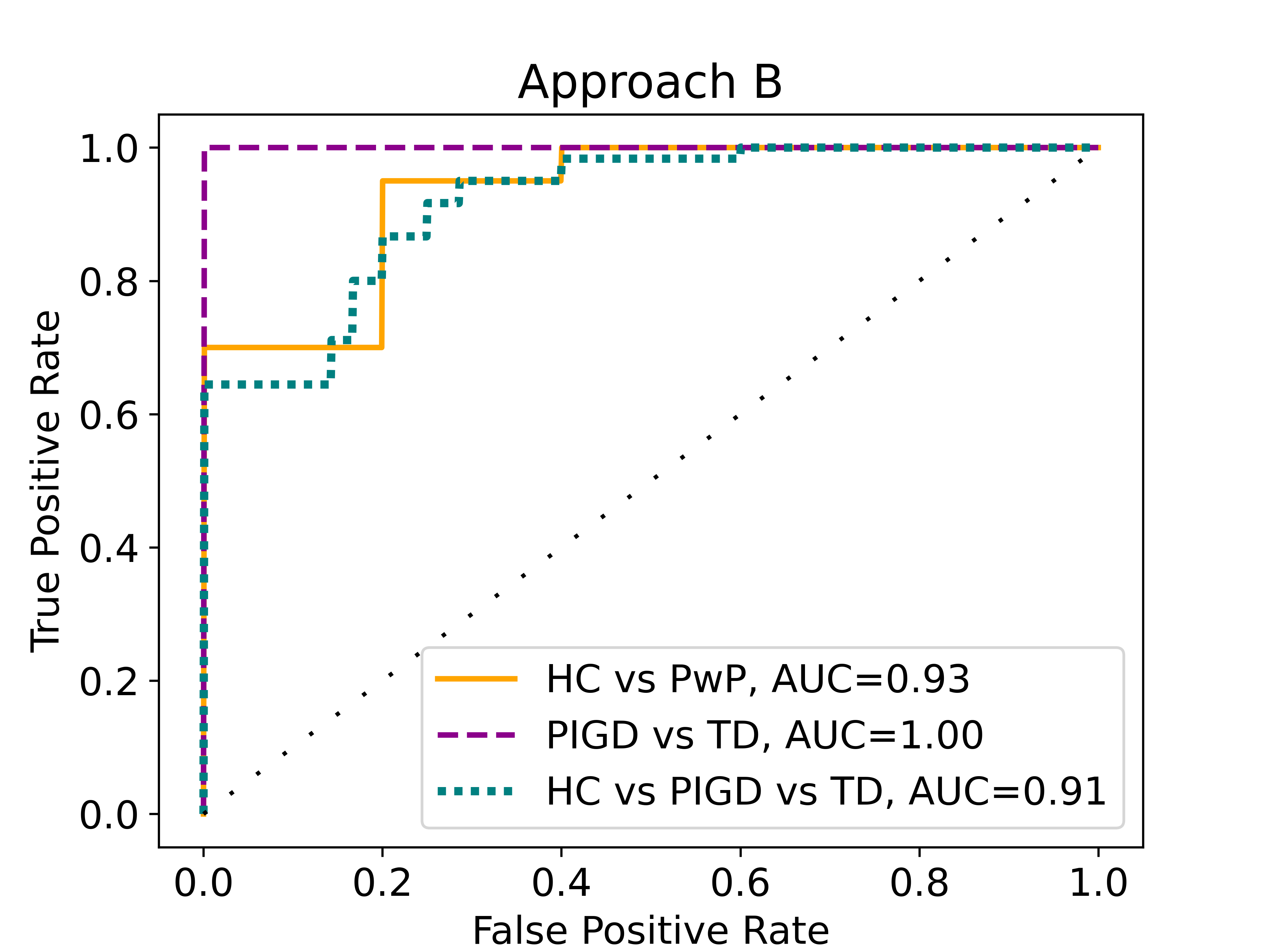}
    \caption{Approach B, using the optimal subset of features: ROC for Healthy Controls (HC) vs People with Parkinson's disease (PwP) in solid orange. ROC for Postural Instability and Gait Difficulty (PIGD) vs Tremor Dominant (TD) in dashed purple. ROC for HC vs PIGD vs TD in dotted green. Chance level is shown in loosely dashed black.}
    \label{fig:fig5}
\end{figure}

\input{tables-figures/tabell5}

%% file: tables-figures/tabell1.tex
\begin{table}[!h]
\centering
\caption{\textbf{Mean volume and mean qMRI values in ROIs.}\\ The mean ROI volumes and the mean R1, R2*, and QSM values of the ROIs for the three classes are presented. Parameter means are reported together with the class‑specific standard deviation of the means, while volumes are presented as mean $\pm$ standard deviation.}
\begin{tabular}{lcccccc}
\toprule
\textbf{ROI} & \textbf{Segmentation method} & \textbf{Volume 10$^{\mathbf{3}}$ [mm$^{\mathbf{3}}$/L]} & \textbf{R1 [s$^{\mathbf{-1}}$]} & \textbf{R2* [s$^{\mathbf{-1}}$]} & \textbf{QSM [ppb]}\\
\midrule
\textit{HC}\super{a} &  &  &  &  & \\
SN\super{b} & \nn{}               & 0.76 $\pm$ 0.09  & 0.71 $\pm$ 0.17 & 77 $\pm$ 23 & 78 $\pm$ 36\\
RN\super{c} & \nn{}               & 0.38 $\pm$ 0.05  & 0.68 $\pm$ 0.17 & 73 $\pm$ 19 & 68 $\pm$ 35 \\
STN\super{d} & \nn{}              & 0.15 $\pm$ 0.03  & 0.71 $\pm$ 0.14 & 68 $\pm$ 19 & 68 $\pm$ 29 \\
Lateral Ventricles & FSV\super{e} & 18.8 $\pm$ 9.81  & 0.30 $\pm$ 0.07 & 10 $\pm$ 16 &  7 $\pm$ 26 \\
Thalamus & FSV           & 9.81 $\pm$ 0.79  & 0.64 $\pm$ 0.07 & 40 $\pm$ 13 &  0 $\pm$ 28 \\
Caudate & FSV            & 4.88 $\pm$ 0.36  & 0.59 $\pm$ 0.05 & 46 $\pm$ 18 & 22 $\pm$ 31 \\
Putamen & FSV            & 6.70 $\pm$ 0.49  & 0.65 $\pm$ 0.04 & 56 $\pm$ 21 & 14 $\pm$ 41 \\
Hippocampus & FSV        & 5.38 $\pm$ 0.43  & 0.57 $\pm$ 0.05 & 34 $\pm$ 16 & -1 $\pm$ 30 \\
Amygdala & FSV           & 2.24 $\pm$ 0.18  & 0.57 $\pm$ 0.04 & 33 $\pm$ 14 & -6 $\pm$ 31 \\
\addlinespace[0.3cm]
\textit{PIGD}\super{f} &  &  &  & &  \\
SN & \nn{}               & 0.72 $\pm$ 0.08 & 0.78 $\pm$ 0.19 & 79 $\pm$ 24 & 83 $\pm$ 40 \\
RN & \nn{}               & 0.33 $\pm$ 0.03 & 0.74 $\pm$ 0.16 & 77 $\pm$ 20 & 74 $\pm$ 36 \\
STN & \nn{}              & 0.14 $\pm$ 0.02 & 0.78 $\pm$ 0.17 & 68 $\pm$ 19 & 60 $\pm$ 28 \\
Lateral Ventricles & FSV & 23.5 $\pm$ 11.3 & 0.31 $\pm$ 0.07 & 9.2 $\pm$ 15 & 6 $\pm$ 24 \\
Thalamus & FSV           & 9.39 $\pm$ 0.63 & 0.64 $\pm$ 0.09 & 39 $\pm$ 12 & 1 $\pm$ 27 \\
Caudate & FSV            & 4.70 $\pm$ 0.42 & 0.59 $\pm$ 0.05 & 45 $\pm$ 17 & 19 $\pm$ 31 \\
Putamen & FSV            & 6.08 $\pm$ 0.62 & 0.65 $\pm$ 0.04 & 55 $\pm$ 20 & 12 $\pm$ 38 \\
Hippocampus & FSV        & 5.16 $\pm$ 0.37 & 0.58 $\pm$ 0.05 & 35 $\pm$ 16 & 1 $\pm$ 27 \\
Amygdala & FSV           & 2.20 $\pm$ 0.25 & 0.58 $\pm$ 0.04 & 33 $\pm$ 14 & -5 $\pm$ 30 \\
\addlinespace[0.3cm]
\textit{TD}\super{g} &  &  &  &  &  \\
SN & \nn{}               & 0.73 $\pm$ 0.08 & 0.75 $\pm$ 0.17 & 79 $\pm$ 24 & 87 $\pm$ 42 \\
RN & \nn{}               & 0.34 $\pm$ 0.02 & 0.72 $\pm$ 0.17 & 75 $\pm$ 19 & 74 $\pm$ 36 \\
STN & \nn{}              & 0.14 $\pm$ 0.03 & 0.71 $\pm$ 0.13 & 67 $\pm$ 18 & 66 $\pm$ 28 \\
Lateral Ventricles & FSV & 20.4 $\pm$ 10.4 & 0.31 $\pm$ 0.07 & 11 $\pm$ 17 & 6 $\pm$ 26 \\
Thalamus & FSV           & 9.51 $\pm$ 0.83 & 0.65 $\pm$ 0.07 & 40 $\pm$ 13 & 1 $\pm$ 29 \\
Caudate & FSV            & 4.83 $\pm$ 0.66 & 0.60 $\pm$ 0.05 & 47 $\pm$ 18 & 22 $\pm$ 32 \\
Putamen & FSV            & 6.20 $\pm$ 0.54 & 0.65 $\pm$ 0.04 & 55 $\pm$ 19 & 11 $\pm$ 40 \\
Hippocampus & FSV        & 5.27 $\pm$ 0.62 & 0.58 $\pm$ 0.06 & 36 $\pm$ 16 & -1 $\pm$ 29 \\
Amygdala & FSV           & 2.12 $\pm$ 0.30 & 0.58 $\pm$ 0.04 & 33 $\pm$ 13 & -6 $\pm$ 30 \\
\bottomrule
\end{tabular}
\label{tab:mean-std-mapvalues-rois}
\begin{tablenotes}
\item[$^{\rm a}$] HC=Healthy Control
\item[$^{\rm b}$] SN=Substantia Nigra
\item[$^{\rm c}$] RN=Red Nucleus
\item[$^{\rm d}$] STN=Subthalamic Nucleus
\item[$^{\rm e}$] FSV=FastSurferVinn
\item[$^{\rm f}$] PIGD=Postural Instability and Gait Difficulty
\item[$^{\rm g}$] TD=Tremor Dominant
\end{tablenotes}
\end{table}

%% file: tables-figures/tabell2.tex
\begin{table}[!h]
\centering
\caption{\textbf{The most important features for Task 1: classifying Healthy Controls (HC) vs People with Parkinson's disease (PwP).}\\The first three columns are the top features from Approach A for the classifiers where this extraction is possible. The rightmost column contains the features from Approach B.}
\begin{tabular}{lcc|c}
\toprule
\textbf{Random Forest} & \textbf{LogReg\super{a}} & \textbf{XGBoost} & \textbf{Approach B} \\
\midrule
Volume of RN\super{b} & Volume of RN & Mean R1 in RN & Volume of RN\\
Min R1 in amygdala & Max R2* in RN  & Median R1 in RN & Volume of putamen \\
Volume of thalamus & Volume of putamen & Skewness of R1 amygdala & 5-percentile R1 in RN\\
Min R1 in RN & Median QSM in putamen & Median R1 in caudate & Median QSM in hippocampus \\
Max R2* in RN & Mean QSM in putamen & Skewness of R1 putamen & 5-percentile R1 in amygdala \\
Skewness of R1 in amygdala & QSM min in RN & Skewness of QSM in putamen & Volume of thalamus \\
Volume of putamen & Min R2* in caudate & Volume of thalamus & STD QSM in hippocampus \\
\bottomrule
\end{tabular}
\label{tab:hcvspwp_features}
\begin{tablenotes}
\item[$^{\rm a}$] LogReg=Logistic Regression
\item[$^{\rm b}$] RN=Red Nucleus
\end{tablenotes}
\end{table}

%% file: tables-figures/tabell3.tex
\begin{table}[!h]
\centering
\caption{\textbf{The most important features for Task 2: classifying Postural Instability and Gait Difficulty (PIGD) vs Tremor Dominant (TD).}\\The first three columns are the top features from Approach A for the classifiers where this extraction is possible. The rightmost column contains the features from Approach B.}
\begin{tabular}{lcc|c}
\toprule
\textbf{Random Forest} & \textbf{LogReg}\super{a} & \textbf{XGBoost} & \textbf{Approach B} \\
\midrule
Kurtosis of R2* in STN\super{b} & Kurtosis of R2* in STN & STD R1 in amygdala & Kurtosis of QSM in putamen\\
STD R2* in hipp & 5-percentile QSM in hippocampus & Kurtosis of R2* in STN & 5-percentile QSM in hippocampus\\
5-percentile R2* in amygdala & 5-percentile QSM in STN & Kurtosis of QSM in putamen & Kurtosis of R2* in STN\\
5-percentile QSM in SN\super{c} & STD R1 in STN & Skewness of QSM in RN & Max R1 in STN\\
\bottomrule
\end{tabular}
\label{tab:pigdvstd_features}
\begin{tablenotes}
\item[$^{\rm a}$] LogReg=Logistic Regression
\item[$^{\rm b}$] STN=Subthalamic Nucleus
\item[$^{\rm c}$] SN=Substantia Nigra

\end{tablenotes}
\end{table}

%% file: tables-figures/tabell4.tex
\begin{table}[!h]
\centering
\caption{\textbf{The most important features for Task 3: classifying Healthy Controls (HC) vs Postural Instability and Gait Difficulty (PIGD) vs Tremor Dominant (TD).}\\The first three columns are the top features from Approach A for the classifiers where this extraction is possible. The rightmost column contains the features from Approach B.}
\begin{tabular}{lcc|c}
\toprule
\textbf{Random Forest} & \textbf{LogReg}\super{a} & \textbf{XGBoost} & \textbf{Approach B}\\
\midrule
Volume of RN\super{b} & Volume of RN & Mean R1 in amygdala & 5-percentile QSM in hippocampus \\
Min R1 in caudate & Max R2* in putamen & Kurtosis of R2* in STN\super{c} & Volume of RN \\
5-percentile of R1 in RN & Kurtosis of R2* in RN & Kurtosis of R1 in amygdala & 5-percentile QSM in STN \\
STD QSM in hippocampus & Kurtosis of QSM in LV\super{d} & Min R1 in caudate & Min R1 in RN \\
Median R2* in putamen & Skewness of QSM in LV & 5-percentile QSM in hippocampus & Volume of putamen \\
Kurtosis of R2* in putamen & 5-percentile of R1 in thalamus & Volume of RN & Kurtosis of R2* in STN \\
\bottomrule
\end{tabular}
\label{tab:hcvspigdvstd_features}
\begin{tablenotes}
\item[$^{\rm a}$] LogReg=Logistic Regression
\item[$^{\rm b}$] RN=Red Nucleus
\item[$^{\rm c}$] STN=Subthalamic Nucleus
\item[$^{\rm d}$] LV=Lateral Ventricles
\end{tablenotes}
\end{table}

%% file: tables-figures/tabell5.tex
\begin{table}[!h]
\centering
\caption{\textbf{Mean classification accuracy and AUC.}\\ Mean accuracy and AUC over the five folds for the three classification tasks. For Approach A the ML classifiers are presented in in row 1-5. Accuracy and AUC for Approach B in the bottom row.}\begin{tabular}{lcccccc}
\toprule
& \multicolumn{2}{c}{\rule{15pt}{0pt} \textbf{HC\super{a} vs PwP\super{b}}\rule{15pt}{0pt} } & \multicolumn{2}{c}{\rule{15pt}{0pt} \textbf{PIGD\super{c} vs TD\super{d}} \rule{15pt}{0pt}} & \multicolumn{2}{c}{\rule{10pt}{0pt} \textbf{HC vs PIGD vs TD}\rule{10pt}{0pt} } \\
\cmidrule(lr){2-3} \cmidrule(lr){4-5} \cmidrule(lr){6-7}
\textbf{Classifier} & \rule{10pt}{0pt} \textbf{ACC\super{e}} & \rule{10pt}{0pt} \textbf{AUC\super{f}} & \rule{7pt}{0pt} \textbf{ACC} & \textbf{AUC} & \rule{12pt}{0pt} \textbf{ACC} & \textbf{AUC} \\
\midrule
\textit{Approach A} &  &  &  &  &  \\
SVM\super{g} & 0.62 & 0.67 & 0.61 & \textbf{0.90} & 0.57 & \textbf{0.66} \\
Random forest & 0.62 & 0.65 & 0.52 & 0.67 & 0.57 & 0.64 \\
LogReg\super{h} & \textbf{0.69} & \textbf{0.73} & \textbf{0.69} & 0.77 & 0.53 & 0.60 \\
XGBoost & 0.64 & 0.68 & 0.61 & 0.67 & 0.54 & 0.65 \\
KNN\super{i} & 0.67 & 0.67 & 0.61 & 0.58 & \textbf{0.62} & 0.65 \\
\addlinespace[0.3cm]
\textit{Approach B} &  &  &  &  &  \\
SVM & 0.82 & 0.93 & 1.00 & 1.00 & 0.73 & 0.91\\
\bottomrule
\end{tabular}
\label{tab:all_classifiers-all_tasks}
\begin{tablenotes}
\item[$^{\rm a}$] HC=Healthy Control
\item[$^{\rm b}$] PwP=People with Parkinson's Disease
\item[$^{\rm c}$] PIGD=Postural Instability and Gait Difficulty
\item[$^{\rm d}$] TD=Tremor Dominant
\item[$^{\rm e}$] ACC=Accuracy
\item[$^{\rm f}$] AUC=Area Under the Curve
\item[$^{\rm g}$] SVM=Support Vector Machine
\item[$^{\rm h}$] LogReg=Logistic Regression
\item[$^{\rm i}$] KNN=K-Nearest-Neighbours
\end{tablenotes}
\end{table}

%% file: sections/discussion.tex
\section{Discussion}
This study evaluated two approaches for classification of HC vs PwP , PIGD vs TD, and HC vs PIGD vs TD. PIGD vs TD was the most successful task in both approaches, although all tasks performed satisfactorily in Approach B. This indicates the heterogeneous characteristics of PD, where qMRI-values are more different between subtypes of PwP than between PwP and HC.

Desai et al. trained a neural network for differentiating PD from HC \cite{desai_enhancing_2024}. Similarly, Zubair et al. trained a neural network on T1w images to classify individuals as HC, prodromal PD or PD \cite{zubair_classification_2024}. Although well-performing, the DL studies use different inputs and problem definitions; in contrast, this present study focuses on qMRI-derived, anatomically interpretable features from specific nuclei to test whether tissue-property signatures can support diagnosis and motor-phenotype stratification.

Gu et al. classified PwP as PIGD or non-PIGD using resting-state fMRI, T1w, and diffusion tensor imaging \cite{gu_automatic_2016}. Hosseini et al. classified PIGD and TD using T1w images \cite{hosseini_cross-regional_2025}. Cheng et al. classified HC, PIGD and TD mutually, using structural images \cite{cheng_explainable_2025}. However, one should be careful when interpreting functional MRI from PwPs, and therefore this present study evaluated the classification performance using only structural MRI.

\subsection{Curse of Dimensionality} 
Approach B, extracting the most relevant features, yielded a significantly better result than letting the classifiers weigh the features. The classifiers should all during fitting, be able to find features that are useful in predicting correct classes, and some of them are even said to work well for cases where number-of-features $>$ number-of-samples \cite{burges_tutorial_1998, huang_parameter_2016}. The divergence between models is consistent with instability in high-dimensional, small-sample settings (p$\gg$n), where correlated features and model-dependent regularisation can yield variable feature rankings. This is a known risk, the \textit{Curse of Dimensionality} \cite{berisha_digital_2021}. 

Consistent with previous studies reporting reduced putamen volume in PD, and inspired by the mentioned studies, a simplified benchmark model for which an SVM was trained using putamen volume alone for the HC vs PwP task. This reduced representation outperformed the SVM trained on the full feature vector, supporting the interpretation that high dimensionality relative to sample size can impair generalisation. This finding motivated Approach B. 

\subsection{Feature Weights} 
The literature mainly focuses on QSM value of subcortical nuclei \cite{langkammer_quantitative_2016, guan_regionally_2017, tan_utility_2021, zang_different_2025, shahmaei_evaluation_2019, he_region-specific_2015, barbosa_quantifying_2015, alushaj_midbrain_2025,heim_magnetic_2017}, so the importance of R1 values  was unexpected.
However, the \textit{all-features} classifiers did not prioritise more than half of the same features as the \textit{optimal-subset} classifier. 
Therefore it can be deduced that the classifiers using the full feature vector were unable to find the features that contain the most information about the classes. 

Therefore, the frequent picked ROIs from Approach B (putamen, hippocampus and STN) could be interesting for further research on subtyping PD. The involvement of SN, STN and putamen is biologically plausible given their role in dopaminergic and basal ganglia circuitry and the sensitivity of QSM/R2* to iron-related susceptibility changes. Other studies investigating differences between subtypes of PD found differences related to iron distribution in basal ganglia \cite{zhang_distribution_2023,li_quantitative_2018, thomas_brain_2020}. The contribution of hippocampus and amygdala may reflect non-motor or broader network involvement that differs across phenotypes. 

In the near term, the most plausible clinical value is not diagnosis replacement but cohort stratification—supporting standardised phenotype assignment for observational studies and trial enrichment. Imaging-derived stratification may be particularly relevant for endpoints linked to gait, balance, and progression.

\subsection{Deep Learning} 
Another approach could be to explore DL for this task, rather than classical ML. DL reduces the need for manual feature engineering and selection. As also demonstrated in this present study, the selection of features is a critical step \cite{ahmed_smart_2025, cayir_feature_2018}. Due to the small dataset used in this study, DL would likely result in overfitting. An additional aim of this study was to identify the most influential features for the classification process, which could provide new insights for future research. Although interpretability is a concern, explainability tools (e.g., saliency/Grad-CAM) could help localise discriminative regions, but would still require larger datasets to avoid overfitting. 

Future work could explore another way to leverage DL. Rather than extracting values from ROIs, the whole image could be fed to the network. However, as mentioned, these approaches address different inputs and problem definitions.

\subsection{ROI Definition} 
Unexpectedly, the SN was not frequently listed in the most important features. A possible limitation was the ROI delineation. For example, the area suggested as a possible biomarker for PD is a dorsal subsegment, mostly overlapping with nigrosome-1 (N1) \cite{brammerloh_swallow_2022}. This present study used the whole ROI, and therefore it is possible that the changes in N1 are lost within the larger SN ROI. There is as of writing no established tool for segmenting this area automatically.

\subsection{Limitations} 
There are several limitations of this study. First, all data were acquired at a single 7T site, and it is not clear how well the findings translate to 3T systems and other acquisition-sites. Second, the sample sizes were modest, leading to wide uncertainty and limiting the power. Third, the TD/PIGD classification is less reliable in patients who are already on dopaminergic therapy, which applies to this cohort. Finally, clinical phenotype labels can fluctuate over time, and labels assigned at the time of scanning may not fully capture longitudinal variability in disease expression.

\subsection{Conclusion} 
The data-driven Approach A, using all features to classify individuals as 1) HC or PwP, 2) PIGD or TD, or 3) HC or PIGD or TD, did not yield satisfactory results, and may have suffered from the curse of dimensionality (best AUCs of 0.73, 0.90 and 0.67, respectively). This study highlights the importance of careful feature selection for the classical ML methods applied in this work, which increased the AUCs to 0.93, 1.00 and 0.91 respectively in Approach B: \textit{optimal-subset}. Nevertheless, these results should be subject to thorough validation on a larger dataset. It is possible that the results using the Approach B are artificially high and prone to overfitting; however five-fold cross-validation was applied to limit this.

The conclusion from this study is therefore twofold: 1) a data-driven approach without careful feature selection is not sufficient, and 2) qMRI-derived features may provide useful candidate biomarkers for PD diagnosis support, but more realistically, motor phenotype stratification in research settings, motivating larger external validation and portability studies.

%% file: sections/supplementary.tex
\renewcommand{\thesection}{S.\arabic{section}} 
\renewcommand{\thefigure}{S\arabic{figure}}
\setcounter{section}{0} 
\setcounter{figure}{0}

\begin{center}
    {\LARGE\textbf{Supplementary Material}}
\end{center}
\vspace{0.5em}
\section{Workflow for Approach B}\label{sec:sankey_plot}
\begin{figure}[!h]
    \centering
    \includegraphics[width=\linewidth] {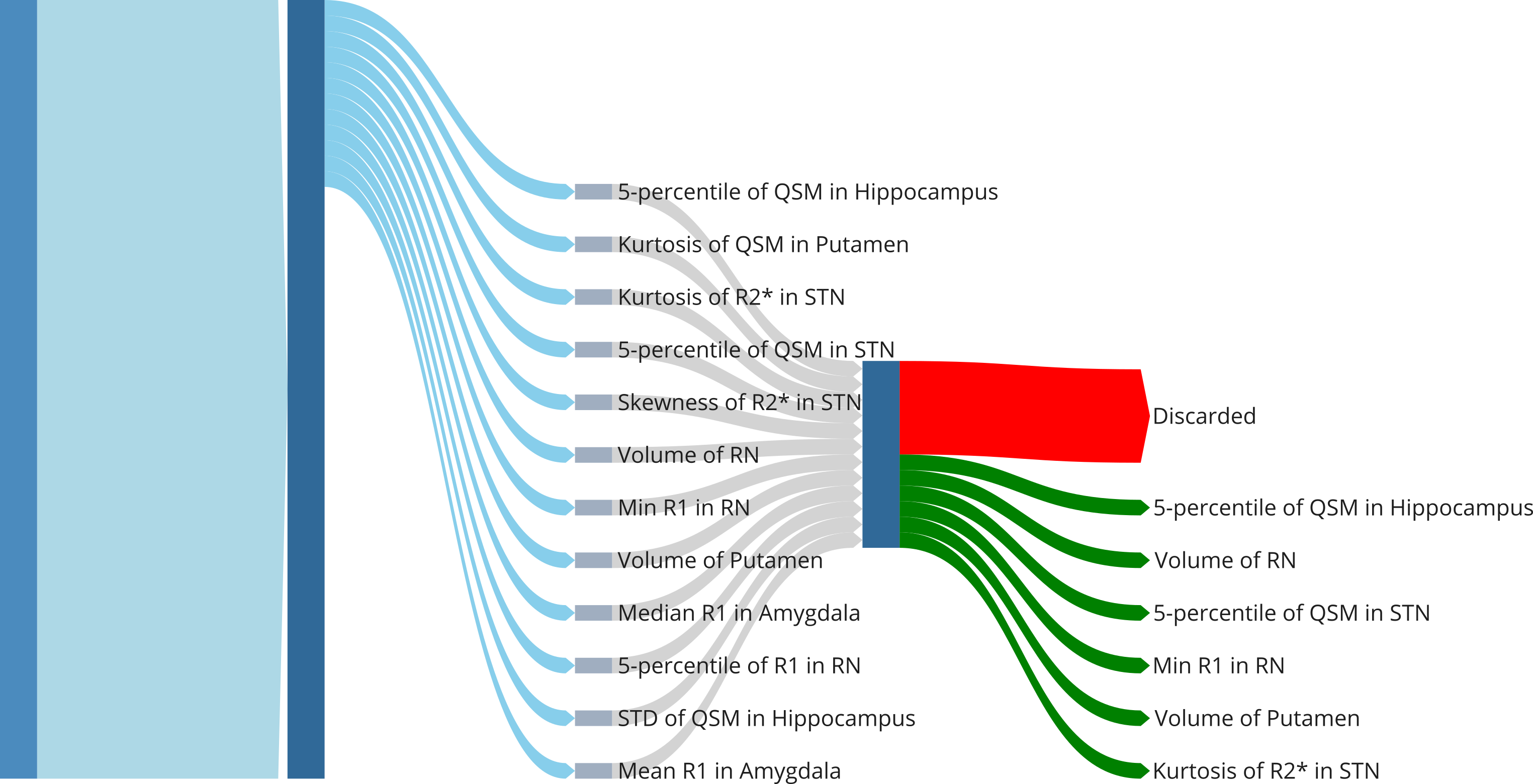} 
    \caption{Sankey plot showing workflow for obtaining the smaller feature pool in the Approach B: \textit{optimal subset}.}
    \label{fig:sankey}
\end{figure}

The workflow for obtaining the features for Approach B, optimal subset, is shown in Figure \ref{fig:sankey} for Task 1: Healthy Controls (HC) vs People with Parkinson's Disease (PwP).

\section{Approach A: All Features}
\subsection{HC vs PwP}
\begin{figure}[!h]
    \centering
    \includegraphics[width=\linewidth]{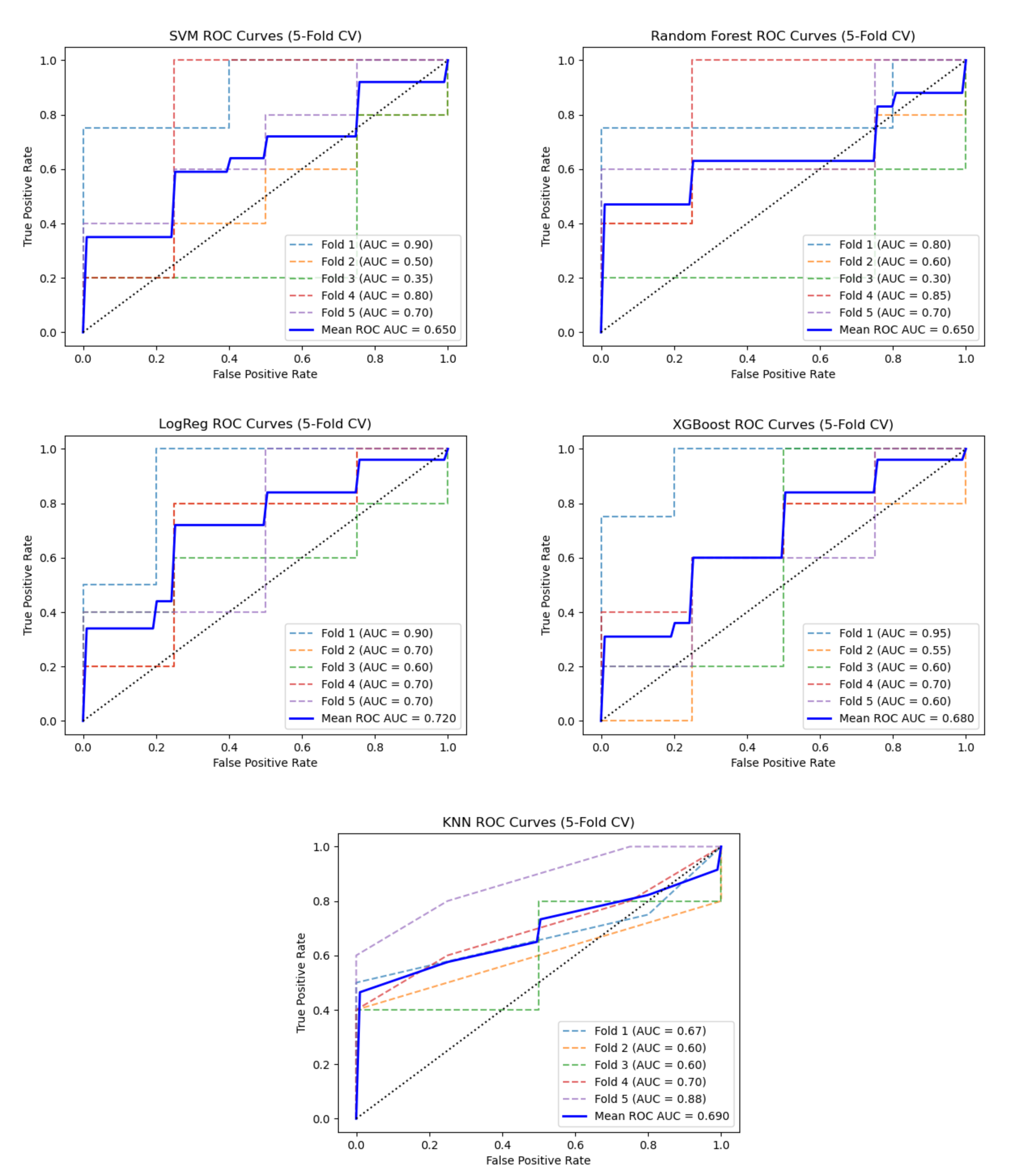}
    \caption{ROC curves for the Approach A: \textit{all features} for HC vs PwP. Each plot shows the ROC curve for the five folds, and a mean ROC curve in thick blue.}
    \label{fig:hcvspwp_rocs}
\end{figure}

In Figure \ref{fig:hcvspwp_rocs} the ROC curves of the classifiers for Task 1: HC vs PwP are shown. As five-fold cross-validation was performed, there are five ROC curves in each plot, and a mean ROC curve.

\subsection{PIGD vs TD}
\begin{figure}[!h]
    \centering
    \includegraphics[width=\linewidth]{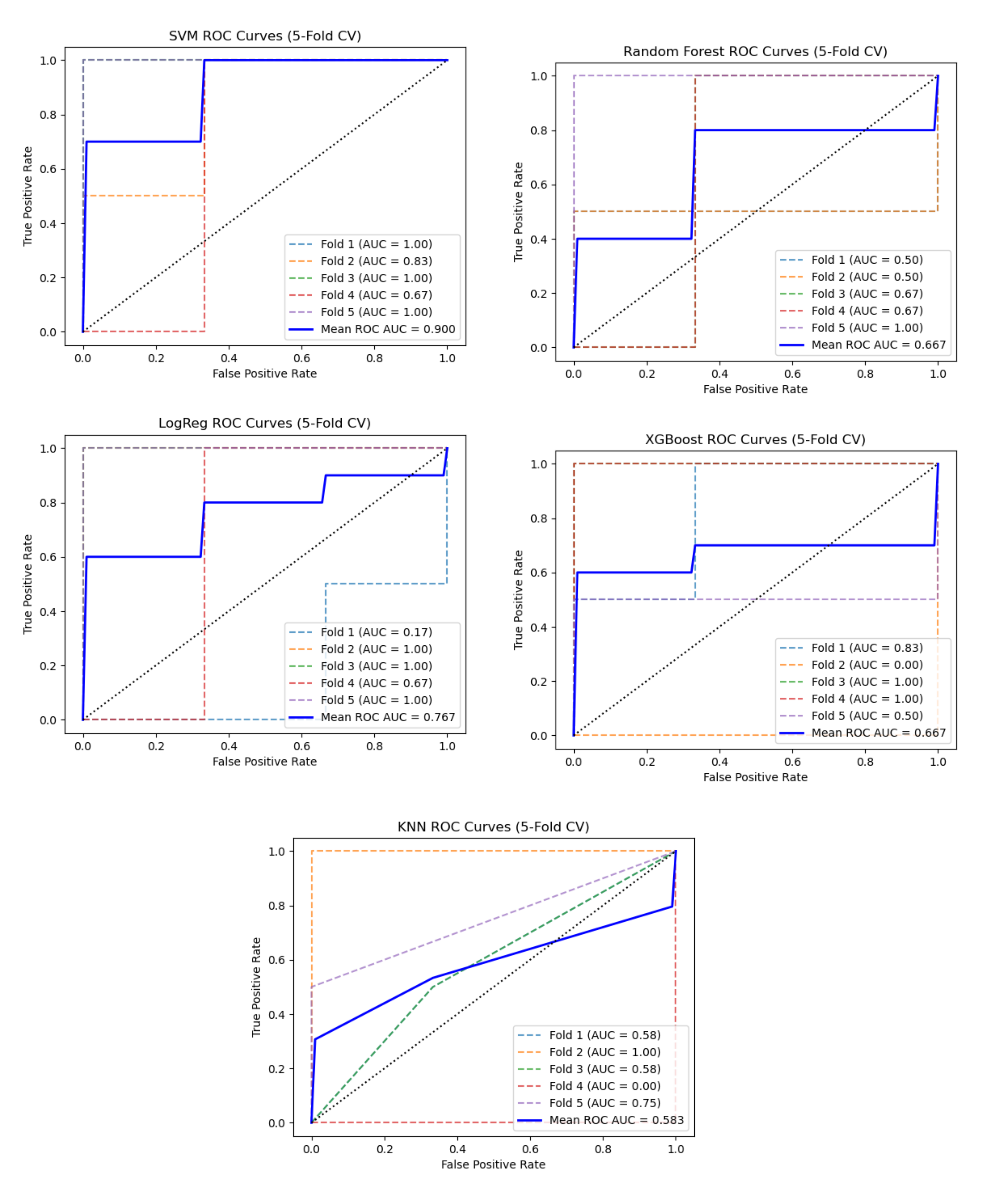}
    \caption{ROC curves for the Approach A: \textit{all features} for PIGD vs TD. Each plot shows the ROC curve for the five folds, and a mean ROC curve in thick blue.}
    \label{fig:pigdvstd_rocs}
\end{figure}

In Figure \ref{fig:pigdvstd_rocs} the ROC curves of the classifiers for Task 2: Postural Instability and Gait Difficulty (PIGD) vs Tremor Dominant (TD) are shown. As five-fold cross-validation was performed, there are five ROC curves in each plot, and a mean ROC curve.

\subsection{HC vs PIGD vs TD}
\begin{figure}[!h]
    \centering
    \includegraphics[width=\linewidth]{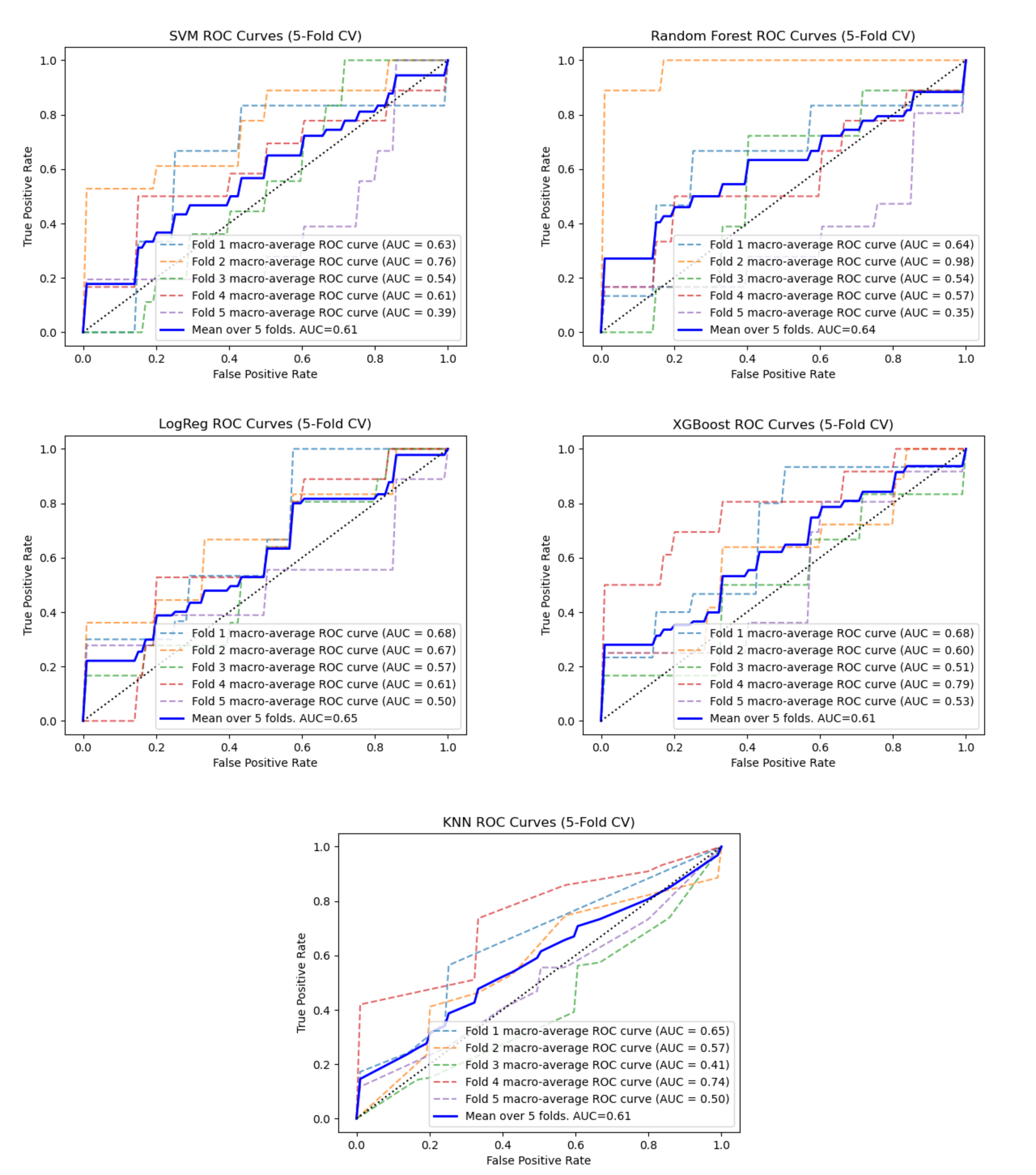}
    \caption{ROC curves for the Approach A: \textit{all features} for HC vs PIGD vs TD. Each plot shows the ROC curve for the five folds, and a mean ROC curve in thick blue.}
    \label{fig:hcvspigdvstd_rocs}
\end{figure}

In Figure \ref{fig:hcvspigdvstd_rocs} the ROC curves of the classifiers for Task 3: HC vs PIGD vs TD are shown. As five-fold cross-validation was performed, there are five ROC curves in each plot, and a mean ROC curve.

\section{Approach B: Optimal Subset}
\begin{figure}[!h]
    \centering
    \includegraphics[width=\linewidth]{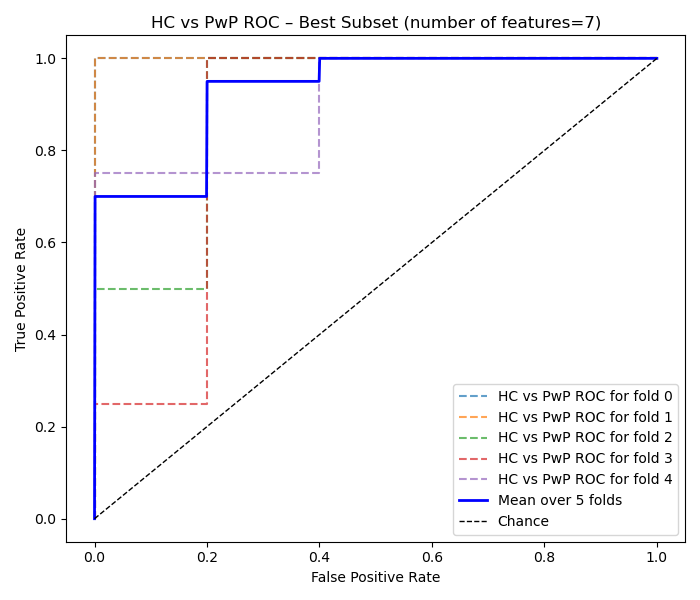}
    \caption{ROC curves for the Approach B: \textit{optimal subset} for HC vs PwP. Each plot shows the ROC curve for the five folds, and a mean ROC curve in thick blue.}
    \label{fig:hcvspwp_optimal_rocs}
\end{figure}

In Figure \ref{fig:hcvspwp_optimal_rocs} the ROC curves of the \textit{optimal-subset} classifiers for Task 1: HC vs PwP are shown. As five-fold cross-validation was performed, there are five ROC curves in each plot, and a mean ROC curve.

\begin{figure}[!h]
    \centering
    \includegraphics[width=\linewidth]{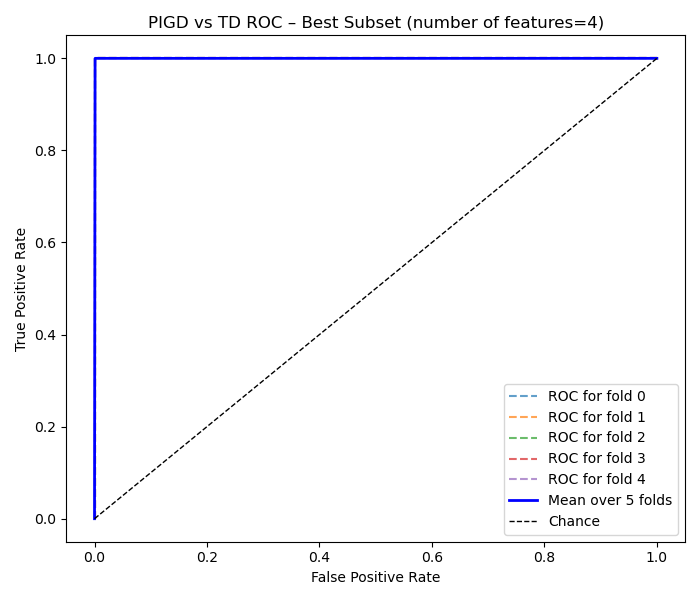}
    \caption{ROC curves for the Approach B: \textit{optimal subset} for PIGD vs TD. Each plot shows the ROC curve for the five folds, and a mean ROC curve in thick blue.}
    \label{fig:pigdvstd_optimal_rocs}
\end{figure}

In Figure \ref{fig:pigdvstd_optimal_rocs} the ROC curves of the \textit{optimal-subset} classifiers for Task 2: PIGD vs TD are shown. As five-fold cross-validation was performed, there are five ROC curves in each plot, and a mean ROC curve.

\begin{figure}[!h]
    \centering
    \includegraphics[width=\linewidth]{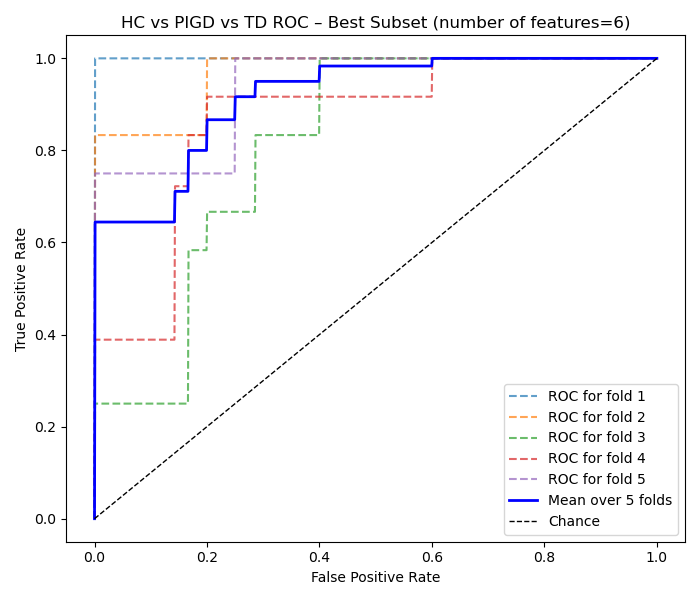}
    \caption{ROC curves for the Approach B: \textit{optimal subset} for HC vs PIGD vs TD. Each plot shows the ROC curve for the five folds, and a mean ROC curve in thick blue.}
    \label{fig:hcvspigdvstd_optimal_rocs}
\end{figure}

In Figure \ref{fig:hcvspigdvstd_optimal_rocs} the ROC curves of the \textit{optimal-subset} classifiers for HC vs PIGD vs TD are shown. As five-fold cross-validation was performed, there are five ROC curves in each plot, and a mean ROC curve.

\subsection{HC vs PwP subset feature pool}\label{sec:hc_pwp_feature_pool}
The 12 features with the highest individual AUC for HC vs PwP classification were min R1 value in RN,  median R1 value in amygdala, 5-percentile R1 value in RN, mean R1 value in amygdala, median R1 value in hippocampus, 5-percentile value of R1 in amygdala, volume of RN, volume of putamen, volume of thalamus, skewness of R2* values in hippocampus, kurtosis of R2* values in hippocampus and STD of QSM values in hippocampus.

\subsection{PIGD vs TD subset feature pool}\label{sec:pigd_td_feature_pool}
The 12 features with the highest individual AUC for PIGD vs TD classification were kurtosis of QSM in putamen, 5-percentile value of QSM in hippocampus, kurtosis of R2* values in STN, 5-percentile value of QSM in STN, skewness of R2* values in STN, STD of QSM values in hippocampus, skewness of QSM values in hippocampus, 5-percentile value of R2* in hippocampus, STD of R1 values in amygdala, mean QSM value in hippocampus, skewness of QSM values in RN and max R1 value in STN.